\PassOptionsToPackage{unicode}{hyperref}
\PassOptionsToPackage{hyphens}{url}
\documentclass[
]{article}
\usepackage{amsmath,amssymb}
\usepackage{iftex}
\ifPDFTeX
  \usepackage[T1]{fontenc}
  \usepackage[utf8]{inputenc}
  \usepackage{textcomp} % provide euro and other symbols
\else % if luatex or xetex
  \usepackage{unicode-math} % this also loads fontspec
  \defaultfontfeatures{Scale=MatchLowercase}
  \defaultfontfeatures[\rmfamily]{Ligatures=TeX,Scale=1}
\fi
\usepackage{lmodern}
\ifPDFTeX\else
\fi
\IfFileExists{upquote.sty}{\usepackage{upquote}}{}
\IfFileExists{microtype.sty}{% use microtype if available
  \usepackage[]{microtype}
  \UseMicrotypeSet[protrusion]{basicmath} % disable protrusion for tt fonts
}{}
\makeatletter
\@ifundefined{KOMAClassName}{% if non-KOMA class
  \IfFileExists{parskip.sty}{%
    \usepackage{parskip}
  }{% else
    \setlength{\parindent}{0pt}
    \setlength{\parskip}{6pt plus 2pt minus 1pt}}
}{% if KOMA class
  \KOMAoptions{parskip=half}}
\makeatother
\usepackage{xcolor}
\usepackage{graphicx}
\usepackage{float}
\usepackage{longtable,booktabs,array}
\usepackage{calc} % for calculating minipage widths
\usepackage{etoolbox}
\makeatletter
\patchcmd\longtable{\par}{\if@noskipsec\mbox{}\fi\par}{}{}
\makeatother
\IfFileExists{footnotehyper.sty}{\usepackage{footnotehyper}}{\usepackage{footnote}}
\makesavenoteenv{longtable}
\providecommand{\tightlist}{%
  \setlength{\itemsep}{0pt}\setlength{\parskip}{0pt}}
\ifLuaTeX
  \usepackage{selnolig}  % disable illegal ligatures
\fi
\usepackage[]{natbib}

\usepackage{bookmark}
\IfFileExists{xurl.sty}{\usepackage{xurl}}{} % add URL line breaks if available
\hypersetup{
  pdftitle={When Should Forecasting Models Be Re-Specified? A Cost-Sensitive Trigger for Adaptive Model-Form Updating},
  pdfauthor={Harrison Katz},
  hidelinks,
  pdfcreator={LaTeX via pandoc}}

\title{When Should Forecasting Models Be Re-Specified? A Cost-Sensitive
Trigger for Adaptive Model-Form Updating}
\author{Harrison Katz\\
\small Forecasting, Data Science, Airbnb, Inc.\\
\small \texttt{harrison.katz@airbnb.com}}

\begin{document}
\maketitle

\section{Abstract}\label{abstract}

Forecasting systems are refreshed at every review period as a matter of routine, and that refresh quietly bundles two operations that need not travel together: estimating parameters and selecting the model form. The second is often unnecessary. This note asks the question that follows. Once a system has settled into a reduced-update policy, when should it break that policy and re-specify the form? We define specification debt as the evidence accumulated against the deployed model form, and we turn it into a cost-sensitive trigger for re-specification. In a closed discrete model space the trigger is a threshold on the negative log posterior probability of the deployed specification. In open production settings the same rule runs on predictive score gaps, stacking weights, or calibrated monitoring diagnostics, and a fixed update frequency turns out to be the special case it produces when evidence against the form accumulates at a constant rate. We then take the rule to all 47,982 monthly M4 series over an exponential smoothing grid, and the results are more cautious than the rule alone would suggest. On these series the choice of model form has little average effect on accuracy. Holding it fixed costs under one percent of accuracy at the benchmark, and on a long-history subsample that cost stays under two percent with no clear rise across horizons, so a cheap fixed cadence matches full updating and the best update frequency is a question of regime rather than a universal constant. At a three-step horizon the trigger is slightly worse than its matched fixed cadence. From six steps out to eighteen the conservative cap-eight trigger carries lower average loss than its matched fixed-eight cadence on a long-history subsample under series-clustered tests. It does not win more often; the gain is consistent with avoiding occasional larger losses when re-specification timing matters. The signal that earns this is the out-of-sample score gap. An in-sample information-criterion analogue of specification debt does not predict realized degradation; it is a complementary diagnostic, modestly linked to the score gap under low-noise, long-window estimation and unreliable under noise. The paper is a decision rule for model-form maintenance and a map of where acting on it pays.

\textbf{Keywords:} forecasting; model update frequency; model form; forecast horizon; exponential smoothing; model monitoring; decision theory; specification debt

\section{1. Introduction}\label{introduction}

Forecasting support systems face a recurring maintenance decision that
is at once statistical and operational. New observations arrive at each
review period, and the system can do one of three things with them:
refresh the parameters under the existing model form, re-select the
model form from a candidate set, or leave the forecast engine alone. In
retail, finance, and supply-chain planning this decision is not made
once but repeated across thousands or millions of related series.
Updating fully and often is expensive, can inject avoidable forecast
instability, and tends to create planning noise. Updating rarely runs
the opposite risk: it can miss structural change and let systematic
forecast errors persist.

The most direct empirical motivation for this note comes from Spiliotis
and Petropoulos \citeyearpar{SpiliotisPetropoulos2024}. They study fixed
model-form update frequencies for univariate forecasting models on the
M4 monthly data and the M5 daily retail data, and their design carefully
separates complete model-form specification from the re-estimation of
smoothing parameters and initial states. What they find is that updating
the model form less often than every review period can retain similar or
even better accuracy while cutting computational time substantially.
They report a second effect as well: less frequent model-form updating
can improve forecast stability, which matters directly for production,
replenishment, and budgeting systems.

That finding reframes the practical question. If a full model-form
update is unnecessary at every review period, the open question is no
longer simply how often to update on average, but when a particular
deployed model form should be revisited. A stable series can tolerate a
long gap between model-form searches. A series that breaks structurally,
picks up a new seasonal pattern, or becomes sensitive to a newly
important event may need attention sooner. Fixed schedules answer the
average-cadence question; the trigger proposed here answers the
series-specific and time-specific one.

This note makes three contributions. The first is a definition:
specification debt is the evidence accumulated against the deployed model
form, which in a closed discrete model space equals the negative log
posterior probability assigned to the deployed specification. From this
we derive a cost-sensitive decision rule that re-specifies the model form
when the expected avoidable loss from keeping the current form exceeds
the cost of re-specification. The rule in turn nests fixed update
frequencies, which are recovered when evidence against the deployed form
accumulates at a constant rate. Adaptive updating is therefore not a
rejection of fixed-frequency updating but a generalization of it, one
that lets the rate of evidence accumulation vary across series, time,
regimes, and business contexts.

We are deliberately modest about what posterior model probabilities can
do in production forecasting. In a small candidate set, say an
exponential smoothing family with a finite collection of model forms,
Bayesian model comparison has a clean interpretation. Most production
systems are not like this. Their model space is open, candidate
specifications are discovered over time, and forecasting teams routinely
compare conventional time series models, gradient boosted trees, neural
networks, hierarchical specifications, and manually adjusted baselines.
In that setting a posterior model probability is often less useful than
a predictive score gap, a stacking weight, or a calibrated monitoring
diagnostic. For that reason the rule we propose is decision-theoretic,
and it is not tied to any single estimator of model evidence.

One word on the empirical study before the theory, because it is more
sober than the rule alone might suggest. We ran the trigger on all 47,982
monthly M4 series rather than a sample, and the scale matters. At the
benchmark horizon the trigger does not beat a well-chosen fixed cadence.
The reason is not subtle: changing the selected model form has little
average effect on these series, a cheap fixed schedule already captures almost all of the available
accuracy, and that leaves little for any re-specification policy to win.
The trigger earns its place in a narrower setting. It
carries lower average loss than a matched fixed cadence once the horizon is long enough for a
mistimed switch to register in the forecast, and it acts on a signal that
actually tracks degradation, the out-of-sample score gap, where the
in-sample information-criterion analogue of specification debt does not.
So the paper is as much a map of when re-specification matters as a method
for doing it.

\section{2. Related work}\label{related-work}

\subsection{2.1 Model update frequency and computational
cost}\label{model-update-frequency-and-computational-cost}

Forecast accuracy has long been the focus of the forecasting literature,
but the cost of producing forecasts now competes with it for attention.
The pressure is sharpest in large-scale retail and marketplace settings,
where estimation may be repeated across many series, locations,
products, and forecast origins
\citep{Seaman2018, PetropoulosGrushkaCockayneSiemsenSpiliotis2025}.
Spiliotis and Petropoulos \citeyearpar{SpiliotisPetropoulos2024} take up
the update frequency of univariate forecasting models directly. Their
experiments span the full range from complete model-form updating at
every review period to no updating at all, with intermediate scenarios
that update smoothing parameters, initial states, or both while the
model form stays fixed, and they evaluate accuracy, computational time,
and forecast instability.

These results belong to a broader cost-aware program in forecasting.
Nikolopoulos and Petropoulos \citeyearpar{NikolopoulosPetropoulos2018}
argue that some suboptimality is acceptable when the computational
savings are material. Yardley and Petropoulos
\citeyearpar{YardleyPetropoulos2021} push for moving past error measures
toward the utility and cost of forecasts. Petropoulos et al.
\citeyearpar{PetropoulosApilettiAssimakopoulosEtAl2022} stress that
forecasting practice is not only modeling but also evaluation,
implementation, and organizational use. This note contributes a formal
trigger to that literature, treating model-form updating as an action
that carries both an expected benefit and an implementation cost.

Closest to the decision-theoretic framing used here, \citet{Katz2026PosteriorLearningDebt} casts retraining for Bayesian prediction systems as a cost-sensitive predictive-regret decision driven by posterior learning debt, the divergence between a reference shadow posterior and the deployed frozen posterior. That paper asks when to refresh a deployed posterior while the model form is held fixed. This note moves the same maintenance decision up a level, from refreshing parameters and posteriors to re-specifying the model form itself.

Recent work on global forecasting models points the same way. Zanotti
\citeyearpar{Zanotti2025} studies retraining frequency for global retail
demand models and finds that retraining less often can preserve accuracy
while reducing computational cost. If anything, this makes adaptive
updating more important rather than less. Once full updating at every
review period is no longer the default, a forecasting system needs a
principled rule for deciding when to break from the reduced-update
cadence.

\subsection{2.2 Rolling-origin evaluation and forecast
comparison}\label{rolling-origin-evaluation-and-forecast-comparison}

Rolling-origin evaluation is still a central tool for assessing time
series forecasting methods \citep{Tashman2000, BergmeirBenitez2012}. It
is the natural choice when the question is not which model forecasts well
in a single split, but how a model behaves as new observations arrive
and its training window moves. The monthly benchmark for the
illustration below comes from the M4 competition
\citep{MakridakisSpiliotisAssimakopoulos2020}. The matching retail
benchmark for large-scale daily demand forecasting comes from the M5
competition, whose background and implementation are described by
\citet{MakridakisSpiliotisAssimakopoulos2022M5background} and whose
accuracy results are reported by
\citet{MakridakisSpiliotisAssimakopoulos2022M5accuracy}.

Forecast comparison methods matter for model monitoring as well. The
Diebold-Mariano framework compares expected predictive accuracy across
competing forecasts \citep{DieboldMariano1995}, proper scoring rules
give calibrated comparisons of probabilistic forecasts
\citep{GneitingRaftery2007}, and calibration and sharpness diagnostics
check whether predictive distributions behave as they should
\citep{GneitingBalabdaouiRaftery2007}, with probability integral
transform diagnostics able to flag distributional misspecification
\citep{DieboldGuntherTay1998}. The rule proposed below can take any of
these diagnostics as inputs. What it adds is the missing action rule:
evidence against the deployed model form should trigger re-specification
only when the expected avoided loss exceeds the cost of acting.

The threshold formulation also connects to sequential monitoring and
change-detection, including sequential probability ratio tests and
CUSUM-type stopping rules \citep{Wald1945,Page1954,BassevilleNikiforov1993}.
The difference lies in the action and cost structure. The stopping time
here is not meant only to detect a distributional change; it is meant to
decide whether re-specifying the model form is worth the cost. A fully
dynamic treatment would pose re-specification as a sequential decision
problem with state-dependent future value, but this note keeps to a
one-step expected-loss rule for transparency and ease of implementation.

\subsection{2.3 Model-form selection and open model
spaces}\label{model-form-selection-and-open-model-spaces}

Exponential smoothing makes a useful motivating case, since it comes with
an established family of interpretable model forms
\citep{Gardner1985, Gardner2006, HyndmanKoehlerSnyderGrose2002, HyndmanKoehlerOrdSnyder2008},
and automatic selection methods for exponential smoothing and ARIMA
models have a long history in forecasting software
\citep{HyndmanKhandakar2008}. Model-form selection in production, though,
rarely happens in a genuinely closed model space. Teams add new
regressors, change seasonal structures, introduce hierarchy, swap loss
functions, alter how outliers are treated, or replace a local statistical
model with a global machine learning model.

This is the distinction between closed and open model comparison
\citep{BernardoSmith1994}. Bayes factors are well defined and useful in
finite or countable candidate spaces with specified priors
\citep{KassRaftery1995}, but they can become fragile once the predictive
setting is open. Stacking takes a different route, targeting predictive
utility by choosing weights that optimize expected log predictive density
\citep{YaoVehtariSimpsonGelman2018}. We treat these as different
operational summaries of the evidence against the deployed specification,
not as interchangeable estimators of one universal quantity.

\section{3. Setup}\label{setup}

Consider a forecasting system observed at review periods
\(t = 1,2,\ldots\). At each one the system holds data \(D_t\) and must
issue forecasts over a horizon \(h\). Write \(M\) for a model form, which
might be an ETS component structure, an ARIMA order, a regression
specification with a particular set of event indicators, or a global
machine learning architecture, and write \(\theta_M\) for the parameters
that go with that form. At time \(t\) the deployed system carries a model
form \(M_t^{dep}\) together with a rule for updating its parameters.

The distinction between updating parameters and updating the model form
is the crux of the problem. A parameter update re-estimates
\(\theta_{M_t^{dep}}\) under the current form, whereas a model-form
update replaces \(M_t^{dep}\) with another form drawn from a candidate
set or from an expanded design space. Spiliotis and Petropoulos
\citeyearpar{SpiliotisPetropoulos2024} make this distinction operational
for ETS models by treating model-form update frequency separately from
updates to the smoothing parameters and initial states. The decision we
study is the decision to update the model form.

At each review period the action is

\[
a_t \in \{\text{keep current model form}, \text{re-specify model form}\}.
\]

Neither action is free. Re-specification consumes computation, analyst
attention, engineering time, validation effort, and governance capacity,
and it can create forecast instability of its own. Keeping an inadequate
form is not free either: it can produce systematic forecast losses, such
as underforecasting holidays, missing a structural break, or issuing
intervals that are too narrow during high-volume periods. The decision is
therefore a comparison between the cost of acting and the expected loss
from not acting.

\section{4. Specification debt}\label{specification-debt}

Start with the case where the model space is finite or countable and
fixed. Let \(\mathcal{M}\) be the candidate set, and let

\[
\pi_t(M) = P(M \mid D_t), \quad M \in \mathcal{M}
\]

be the posterior model probabilities after data through review period
\(t\). The deployed system puts all of its operational weight on a single
specification \(M_t^{dep}\). In this closed discrete setting we define
specification debt as

\[
D_t^{spec} = -\log \pi_t(M_t^{dep}).
\]

Equivalently,
\[
D_t^{spec} = \mathrm{KL}(\delta_{M_t^{dep}}\|\pi_t),
\]
the finite KL direction from the deployed point mass to the posterior over
model forms. The reverse direction \(\mathrm{KL}(\pi_t\|\delta_{M_t^{dep}})\)
is typically infinite, since any non-deployed form with positive posterior
mass makes it blow up. Specification debt stays small while the data still
place high posterior probability on the deployed form and grows as that
form loses support. All logarithms are natural.

The definition earns its keep by being finite, monotone in the posterior
evidence against the deployed form, and easy to read. It is not a claim
that production forecasting is usually M-closed. When the model space is
open, exact posterior model probabilities may be unavailable or simply
unhelpful, and the operational object becomes a calibrated surrogate for
the probability, or the expected loss, that the deployed form is
actionably wrong. Predictive score gaps, stacking weights, residual
diagnostics, and calibration diagnostics can all play that part.

The phrase that does the work is actionably wrong. A model can have low
posterior probability as a literal data-generating process and still be
perfectly adequate for the business decision at hand. The reverse also
happens: a model can look fine on average yet be actionably wrong in
exactly the periods that matter most. Specification debt is therefore not
the same as model-selection uncertainty. It is evidence against the
deployed form, measured relative to the decision that form supports.

\section{5. A cost-sensitive trigger}\label{a-cost-sensitive-trigger}

Let \(q_t\) be the posterior or calibrated operational probability that
the deployed model form is actionably wrong. Let \(K\) be the number of
future decision periods over which misspecification costs accumulate. Let
\(c_B\) be the expected per-period incremental loss given that the
deployed form is actionably wrong, and let \(c_R\) be the cost of
re-specification. That cost \(c_R\) folds in computational time, cloud
spend, engineering work, analyst time, validation, governance, and the
cost of any added forecast instability.

The general rule is an expected-loss comparison: re-specify when

\[
E_t\!\left[\sum_{j=1}^{K} \bigl\{L_{t+j}(\text{keep}) - L_{t+j}(\text{re-specify})\bigr\}\right] > c_R,
\]

the expected loss avoided over the decision window against the cost of
acting. A binary approximation, in which an actionably wrong form costs
\(c_B\) per period with probability \(q_t\) and nothing otherwise, reduces
this to

\[
\text{re-specify if } q_t K c_B > c_R.
\]

Put plainly, the system should re-specify when the expected avoidable
loss from keeping the current model form exceeds the cost of changing it.
Two features of the rule are worth drawing out. It will not fire just
because a challenger model edges out the incumbent in a backtest by a
little, and it will not sit quietly through a recurring structural failure
merely because the routine parameter updates are still running.

In the closed discrete simplification, write \(p_t = \pi_t(M_t^{dep})\).
The equality

\[
q_t = 1 - p_t = 1 - \exp(-D_t^{spec})
\]

is a deliberately strong special case, since it treats every non-deployed
model form as actionably different from the deployed one. In general the
decision-relevant probability of actionable misspecification is no larger
than \(1-p_t\), because some posterior mass can sit on alternatives that
are statistically distinct but operationally immaterial. Under the equality the threshold below is
exact; in general, substituting the bound gives

\[
1 - \exp(-D_t^{spec}) > \frac{c_R}{Kc_B}.
\]

Provided \(c_R < Kc_B\), this rearranges to

\[
D_t^{spec} > -\log\left(1 - \frac{c_R}{Kc_B}\right).
\]

The closed-form threshold reads cleanly. Cheap re-specification gives a
low threshold, expensive re-specification a high one, and if the cost of
re-specification exceeds the largest avoided loss that is plausible over
the decision window, the rule never fires. Because \(q_t \le 1 - p_t\), failing to cross the threshold rules
re-specification out under the binary-loss approximation, while crossing
it establishes only that acting may be justified, and the decision still
needs a calibrated estimate of actionable loss. The threshold is a
necessary screening condition rather than a sufficient action rule,
which is the role it plays in the two-stage design of Section 7, a cheap
screen in front of the costed decision. The equality
\(q_t = 1 - p_t\) is the special case in which screen and trigger
coincide, and the general rule is the expected-loss inequality.

A score-based implementation swaps \(q_t K c_B\) for an estimated
avoidable score loss. Let \(S(y, F)\) be a proper score, smaller being
better, let \(F_t^{dep}\) be the predictive distribution from the deployed
form, and let \(F_t^{ch}\) be the best challenger the monitoring system
has on hand. A rolling validation estimate of score debt is

\[
\Delta_t = \widehat{E}\{S(Y, F_t^{dep}) - S(Y, F_t^{ch})\}.
\]

The expectation is taken over a rolling validation window and, where it
makes sense, averaged across forecast horizons. For point forecasts \(S\)
can be an absolute or squared scaled error; for probabilistic forecasts it
can be the negative log score, CRPS, or another strictly proper score. The
cost-sensitive score trigger is then

\[
K \cdot \max(\Delta_t,0) \cdot c_S > c_R,
\]

where \(c_S\) converts the scoring-unit difference into decision cost.
When that scale is uncertain, a sensitivity analysis over \(c_R/(Kc_S)\)
tells you more than committing to one universal threshold.

\section{6. Fixed update frequencies as a special
case}\label{fixed-update-frequencies-as-a-special-case}

Fixed-frequency model-form updating falls out of the threshold rule as a
special case. The debt level \(D_t^{spec}\) need not reset when the form
is re-specified, since the newly selected form rarely carries posterior
probability one, so the object that accumulates from zero is the
increment since the last re-specification. Write \(t_0\) for the most
recent re-specification time and define

\[
B_{t_0,k} = -\log \frac{\pi_{t_0+k}(M_{t_0}^{dep})}{\pi_{t_0}(M_{t_0}^{dep})},
\]

the net log change in posterior support for the deployed form since it
was selected, positive when support has eroded, with \(B_{t_0,0} = 0\) by
construction. Let \(b\) be a threshold on this
increment, and suppose that after a model-form update the evidence
accumulates deterministically at rate \(\lambda > 0\) per review period,
\(B_{t_0,k} = \lambda k\). The threshold stopping time is then

\[
\tau_b = \inf\{k: B_{t_0,k} \ge b\}
       = \left\lceil \frac{b}{\lambda} \right\rceil.
\]

So a fixed update frequency is exactly the threshold rule under constant
evidence accumulation, which gives a formal nesting of fixed-frequency
updating inside adaptive updating. A fixed schedule is reasonable when the
evidence accumulation rate is homogeneous across time and series; an
adaptive schedule earns its place when the rate varies.

The heterogeneous case is the one that shows up in practice. Some series
stay stable for long stretches. Others lurch through abrupt changes in
level, trend, seasonality, event sensitivity, intermittency, or variance.
No single fixed update frequency can be optimal across all of them, unless
the cost of hunting for heterogeneity outweighs the benefit of adapting to
it. This squares with what Spiliotis and Petropoulos
\citeyearpar{SpiliotisPetropoulos2024} find: a fixed-frequency design
already shows that full model-form search at every review period is often
unnecessary. The adaptive extension asks the next question, which is which
series ought to be exempted from a lower-cost default schedule.

The same reasoning carries over to the update scenarios in their study.
The N scenario leaves both the model form and the parameters unchanged
between model-form updates, while the SP, IS, and IS-SP scenarios hold the
form fixed and update selected parameters. In the language of this note
these are all reduced-cost maintenance policies, and an adaptive trigger
can sit on top of any of them. A system can refresh initial states or
smoothing parameters as a matter of routine and re-specify the model form
only when the evidence against the current form crosses a cost-sensitive
threshold.

\clearpage
\section{7. Operational estimators of specification
debt}\label{operational-estimators-of-specification-debt}

Different forecasting environments support different measures of evidence,
and the rule does not ask every system to compute posterior model
probabilities. Table \ref{tab:operational-estimators} collects five common
operational estimators of specification debt and shows how each enters the
trigger.

\refstepcounter{table}\label{tab:operational-estimators}
\begin{center}
\textbf{Table \thetable: Operational estimators of specification debt.}

\small
\begin{tabular}{p{0.22\linewidth}p{0.22\linewidth}p{0.25\linewidth}p{0.23\linewidth}}
\toprule
Setting & Evidence measure & Interpretation & Use in the trigger \\
\midrule
Closed finite candidate set & Posterior model probability or Bayes factor & Data support for the deployed model form & Use \(D_t^{spec} = -\log \pi_t(M_t^{dep})\) \\
Predictive model comparison & Rolling predictive score gap & Estimated loss from keeping the deployed form rather than a challenger & Use \(\Delta_t\) in the score-based trigger \\
Forecast combination & Stacking weight & Predictive usefulness of the deployed form in an ensemble & Trigger when the deployed weight collapses and the decision loss is material \\
Probabilistic monitoring & Calibration, PIT, and coverage diagnostics & Distributional failure of the deployed predictive distribution & Trigger targeted redesign of the uncertainty model \\
Business-process monitoring & Error concentration by event, segment, or horizon & Local structural inadequacy & Trigger targeted candidate generation and validation \\
\bottomrule
\end{tabular}
\end{center}
\normalsize

A practical implementation can run in two stages. The first stage is cheap
monitoring, tracking rolling errors, score gaps, coverage, residual
autocorrelation, override frequency, and performance broken out by segment
or event type. The second stage is conditional search: only once the
monitoring statistic crosses a screening threshold does the system run a
fuller model-form comparison. This two-stage design answers the main
computational objection to adaptive updating, which is that if the trigger
itself required fitting all candidate forms at every review period, it
would erase the savings from less frequent model-form updates in the first
place.

Algorithm 1 gives a generic score-triggered version.

\textbf{Algorithm 1. Adaptive model-form updating by score debt}

\begin{enumerate}
\def\labelenumi{\arabic{enumi}.}
\tightlist
\item
  At the initial review period, select a deployed form \(M^{dep}\) from
  a candidate set using the chosen model-selection rule.
\item
  At each review period \(t\), update parameters under \(M^{dep}\)
  according to the routine maintenance policy.
\item
  Compute a low-cost monitoring statistic, such as a rolling score gap,
  coverage failure, or error concentration measure.
\item
  If the monitoring statistic is below the screening threshold, keep
  \(M^{dep}\).
\item
  If the monitoring statistic exceeds the screening threshold, evaluate
  a challenger set on a rolling validation window.
\item
  Estimate avoidable loss from replacing \(M^{dep}\) with the best
  challenger.
\item
  Re-specify only if estimated avoidable loss exceeds re-specification
  cost.
\item
  Record the action, score gap, computational cost, and forecast
  instability for audit and calibration.
\end{enumerate}

The trigger should be calibrated on historical rolling-origin evaluations
rather than fixed once by convention. The calibration is much like
choosing a fixed update frequency, except that it produces a policy
defined by evidence and cost instead of by the calendar alone.

\section{8. Empirical evaluation}\label{empirical-evaluation}

The evaluation uses the monthly subset of the M4 competition data and complements the update-frequency experiments of Spiliotis and Petropoulos \citeyearpar{SpiliotisPetropoulos2024} rather than reproducing them in full. Their study evaluates fixed model-form update frequencies on the M4 monthly data with 36 rolling-origin rounds, horizon three, and frequencies from 1 through 12, reporting accuracy, computational time, and forecast instability. We keep those three dimensions and add capped adaptive policies that re-specify the model form when a score-gap trigger fires or when the deployed form reaches a maximum age. We then go past the single benchmark configuration in two ways: we run every eligible series rather than a sample, and we vary the forecast horizon to ask whether the benchmark result is specific to short-horizon forecasting.

\subsection{8.1 Data and evaluation design}\label{data-and-evaluation-design}

The code merges the M4 monthly training and test files and applies the rolling-origin eligibility filter, which leaves 47,982 monthly series, the usable count reported by Spiliotis and Petropoulos \citeyearpar{SpiliotisPetropoulos2024}. The main evaluation runs every one of these series rather than a sample. Each policy is run over 36 rolling-origin rounds with a fixed training window of 36 observations, seasonal period \(s=12\), and forecast horizon \(h=3\) at the benchmark, producing 1,727,352 forecast-origin records per policy. The 36-observation window follows the M4 monthly setup of Spiliotis and Petropoulos and amounts to three seasonal cycles. Fitting is near-universal: depending on the policy, between 216 and 276 of the 1,727,352 origin fits, at most 0.016 percent, fall back to a seasonal naive forecast when every candidate fit fails, and no origin is dropped or recorded with a missing loss. Relative loss for a policy is the ratio of its grand-mean MASE over all forecast-origin records to the full-update grand mean, whose absolute level is 0.639; relative time and relative instability are built the same way. The MASE scale is the mean absolute seasonal difference on the training window; when that scale falls below \(10^{-12}\) the first-difference scale is used instead, and one is used if both degenerate, so every origin carries a finite loss. sMAPC terms whose adjacent forecasts are both zero are excluded from the average.

Accuracy is measured by MASE, the mean absolute scaled error of Hyndman and Koehler \citeyearpar{HyndmanKoehler2006}. Computational time is the wall-clock seconds spent in model selection, refitting, and monitoring. Forecast instability is measured by sMAPC, the symmetric mean absolute percentage change between adjacent forecast origins, following the instability emphasis of Spiliotis and Petropoulos \citeyearpar{SpiliotisPetropoulos2024}:

\[
\mathrm{sMAPC}_t = \frac{200}{h-1}\sum_{j=1}^{h-1}
\frac{|\widehat y_{t+j\mid t} - \widehat y_{t+j\mid t-1}|}{|\widehat y_{t+j\mid t}| + |\widehat y_{t+j\mid t-1}|}.
\]
All empirical results are reported relative to the full-update benchmark, the policy that selects the model form at every review period.

Two further designs support the horizon analysis. To test whether the benchmark result depends on the three-step horizon, we fix a subsample of 4,000 series long enough to support the longest horizon and evaluate that same subsample at horizons 3, 6, 9, 12, and 18, holding the seasonal period at 12, so that only forecast depth changes across the sweep. And because the policies differ only on the minority of rounds where their re-specification schedules diverge, we compare a pair of policies round by round: we pair the per-origin losses on the same series and the same origin, which removes the large between-series variance that would otherwise swamp a sub-percent difference, and we read the paired difference by sign, by median, and by mean, together with a decomposition of the largest divergences and a check on whether they concentrate in a few series.

\subsection{8.2 Model forms and policies}\label{model-forms-and-policies}

The implementation works with an explicit ETS candidate grid fit through \texttt{forecast::ets}. The grid holds six additive-error forms, ETS(A,N,N), ETS(A,A,N), ETS(A,A$_d$,N), ETS(A,N,A), ETS(A,A,A), and ETS(A,A$_d$,A), and, for strictly positive series, nine multiplicative forms, ETS(A,N,M), ETS(A,A,M), ETS(A,A$_d$,M), ETS(M,N,N), ETS(M,A,N), ETS(M,A$_d$,N), ETS(M,N,M), ETS(M,A,M), and ETS(M,A$_d$,M), fifteen candidates in all and six otherwise, the same kind of finite ETS space used in the update-frequency literature. Multiplicative trends are excluded, no Box-Cox transformation is applied, and \texttt{forecast::ets} runs with its default admissibility bounds. Scheduled selection, at initialization and at every fixed-cadence or cap-driven re-specification, takes the minimum-AICc candidate on the full current training window; a candidate whose fit fails is dropped from that comparison, and if every candidate fails the origin falls back to a seasonal naive forecast, the rule behind the fallback counts in Section 8.1. The candidates cover level-only, trended, damped-trend, seasonal, and damped-trend seasonal variants. Under the reduced-update policies the selected form is carried forward until the next re-specification.

The policies are:

\begin{itemize}
\item \textbf{Full update}: select the model form and fit parameters at every review period.
\item \textbf{Fixed-\(f\) update}: select the model form every \(f\) review periods for \(f=2,\ldots,12\); between model-form searches, keep the form fixed and refit parameters.
\item \textbf{Parameter-only update}: select the initial model form once and refit parameters under that form at each review period.
\item \textbf{Capped adaptive update}: keep the form fixed unless a rolling score-gap trigger fires or the form reaches a maximum age. The main run uses caps of 8 and 12 review periods, thresholds \(\tau \in \{0.03,0.05,0.10,0.20,0.40,0.80\}\), a monitoring window of 12 observations, and monitoring every 6 review periods.
\end{itemize}

The empirical threshold \(\tau\) is the score-unit version of the cost ratio \(c_R/(Kc_S)\) from Section 5. A lower \(\tau\) corresponds to cheaper re-specification, a larger decision cost on avoidable score loss, or both, while a higher \(\tau\) corresponds to more expensive re-specification or less sensitivity to the observed score gap. The capped policy nests the fixed schedules: if the score-gap trigger never fires the cap sets the update frequency, and if evidence accumulates sooner the form is re-specified ahead of the cap. This is the empirical counterpart of the formal nesting in Section 6.

The trigger's operational timeline at a monitored origin, every sixth review period, runs as follows. The 36-observation training window splits into a base of 24 observations and a validation tail of 12, the monitoring window. The deployed form is refit on the base and scored on the tail, a 12-step MASE whose scale comes from the base. Every candidate in the grid is then fit on the base and scored on the same tail, and the challenger is the candidate with the lowest validation loss. The score gap is the deployed validation loss minus the challenger's, and the trigger fires when the gap exceeds \(\tau\) and the challenger differs from the deployed form. A score-triggered re-specification deploys the validation winner, whereas scheduled and cap-driven re-specifications select by AICc on the full window. The cap counts age from the most recent re-specification of either kind, and the capped policy shares its initial selection and calendar with its matched fixed cadence, so the two coincide on every round where the trigger has not fired. The validation tail lies strictly before the forecast origin, so it never overlaps the evaluation horizon, and all monitoring computation is charged to the policy's time. In this fifteen-form grid the validation comparison itself serves as the monitored score signal at the six-period review points; a larger production space would need the cheaper first-stage screen of Algorithm 1 in front of the challenger search.

Because the challenger is selected and its gap estimated on the same twelve observations, the gap is a selected comparison and is biased upward, so \(\tau\) is calibrated to the selected-challenger procedure rather than to an unbiased loss difference; that selection is one reason the lowest thresholds over-fire in Section 8.3 and three-step firings chase noise in Section 8.4. And because the monitoring window is twelve observations at every forecast horizon, the gap's scale does not change across the horizon sweep, so a single \(\tau\) is comparable at all depths. The horizon sweep uses a reduced grid for cost, with caps at 8, threshold 0.8, and fixed cadences at 8 and 12, which is enough to read both the cost of a stale form and the matched comparison of the trigger against its own cadence at each horizon. The sweep fixes the adaptive policy at cap eight and threshold 0.8, the least trigger-sensitive threshold in the grid and one of the two strongest adaptive policies on the three-step benchmark, chosen on that benchmark before any longer horizon was run. The sweep is therefore a policy-specific matched comparison rather than a search over the adaptive grid, and we read it as evidence about this conservative trigger, not as post-selection proof that adaptive updating dominates fixed schedules.

\subsection{8.3 The cost of model-form choice}\label{the-cost-of-model-form-choice}

Table \ref{tab:m4-frontier} reports selected policies over all 47,982 series at the benchmark horizon. What the full run shows first is how little separates the sensible policies. Set computational time aside and the fixed cadences sit within a quarter of a percent of full updating on accuracy, top to bottom. The largest gap among the low-cost policies is parameter-only updating, which never changes the model form at all and still costs only 0.87 percent, at 6 percent of the compute. A cheap fixed cadence is closer still: \texttt{fixed\_f8} sits 0.06 percent from full updating at 16.7 percent of the compute. The one exception is the trigger-happy end. Re-specify on the slightest score gap and accuracy falls off, by as much as 3.4 percent at the lowest thresholds. Full updating is matched, not beaten, by schedules that do a small fraction of its work.

\begin{figure}[H]
\centering
\includegraphics[width=0.92\linewidth]{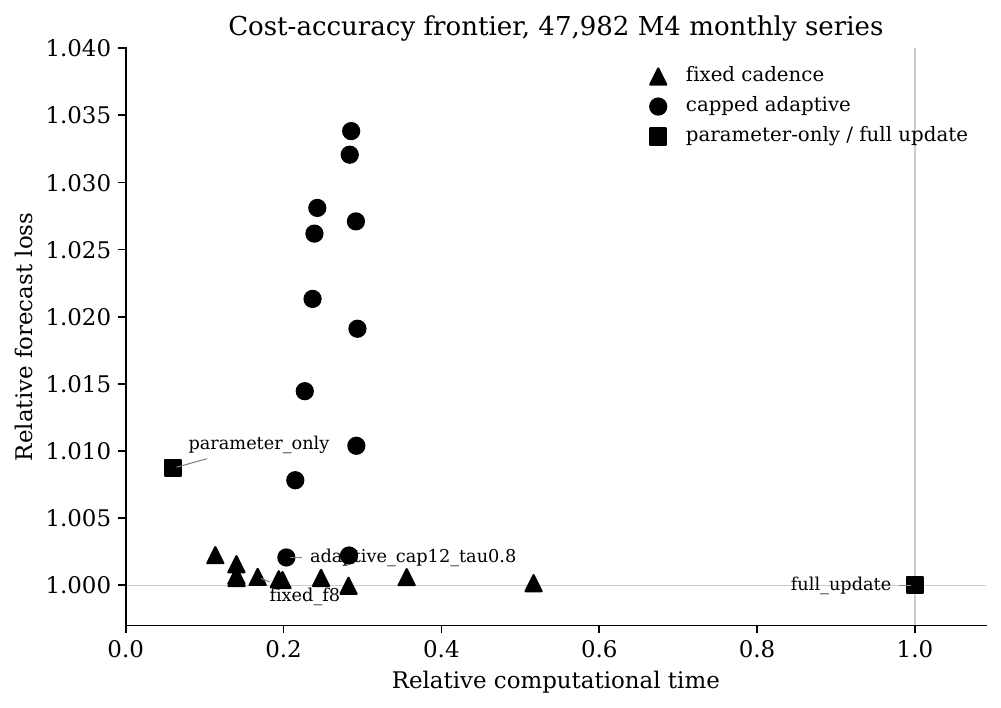}
\caption{Cost-accuracy frontier over all 47,982 M4 monthly series at horizon three. Lower values are better on both axes. The cheap fixed cadences match full updating in accuracy at a fraction of the computational time, and the best capped adaptive policies sit near the frontier without leading it, while the lowest-threshold triggers give up accuracy and rise well above it.}
\label{fig:frontier-full}
\end{figure}

Within that narrow band the fixed cadences, not the adaptive policies, sit at the front. Order the table by accuracy and the fixed schedules come first. The best adaptive policy, the cap-twelve trigger at threshold 0.8, has relative loss 1.0021, and it is dominated outright by cheaper and more accurate fixed schedules such as \texttt{fixed\_f6} at 1.0004 and \texttt{fixed\_f9} at 1.0005. The capped adaptive policy sits just off the frontier rather than leading it. Figure \ref{fig:frontier-full} shows the same ordering against computational time.

\begin{table}[H]
\centering
\caption{Selected policies over all 47,982 M4 monthly series at horizon three. Values are relative to full updating except for re-specifications per series. Policies are ordered by relative loss. At five decimals \texttt{fixed\_f4} is 0.99996, a hair inside full updating and within sampling noise of it.}
\label{tab:m4-frontier}
\begin{tabular}{lrrrr}
\toprule
Policy & Relative loss & Relative time & Relative instability & Re-specifications \\
\midrule
\texttt{full\_update} & 1.0000 & 1.000 & 1.000 & 36.00 \\
\texttt{fixed\_f4} & 1.0000 & 0.282 & 0.947 & 9.00 \\
\texttt{fixed\_f6} & 1.0004 & 0.199 & 0.942 & 6.00 \\
\texttt{fixed\_f9} & 1.0005 & 0.140 & 0.933 & 4.00 \\
\texttt{fixed\_f8} & 1.0006 & 0.167 & 0.941 & 5.00 \\
\texttt{adaptive\_cap12\_tau0.8} & 1.0021 & 0.204 & 0.930 & 3.22 \\
\texttt{adaptive\_cap8\_tau0.8} & 1.0022 & 0.283 & 0.937 & 5.10 \\
\texttt{fixed\_f12} & 1.0022 & 0.114 & 0.938 & 3.00 \\
\texttt{parameter\_only} & 1.0087 & 0.060 & 0.921 & 1.00 \\
\bottomrule
\end{tabular}
\end{table}

What this shows is not about the trigger. It is about magnitude. On these monthly series, changing the selected ETS form has little average effect, a cheap fixed cadence captures almost all of the available accuracy, and there is little headroom left for any re-specification policy to win. The follow-on is that the best update frequency depends on the regime, and controlled experiments on series with injected structural change make that case from the other side. Give a series an abrupt and persistent shift and full re-fitting dominates, even against a fast-firing trigger, because the right move is to rebuild the form at once. Contaminate it with outliers or a transient burst and the rarely-firing policies win instead, on restraint rather than detection, since acting on the contamination costs more than ignoring it. No fixed frequency is right across both (Appendix C.1). That heterogeneity is what the adaptive rule is built to exploit and the clock cannot.

\subsection{8.4 Forecast horizon and the value of timing}\label{forecast-horizon-and-the-value-of-timing}

At the benchmark horizon the capped adaptive trigger does not beat its own matched cadence. Most of the time the two are the same forecast. The cap-eight trigger at threshold 0.8 and \texttt{fixed\_f8} agree on 91.1 percent of rounds, because on those rounds the trigger has not fired and both deploy the same form on the same data; they part only on the remaining 8.9 percent. The raw spread across series dwarfs any difference between the two policies, so we work with within-series differences: aggregate each series' loss difference and test across all 47,982 series. Resolved this way, the difference runs against the trigger. The per-series mean loss difference is $+0.0010$ in favor of the cadence, small but firmly nonzero, with a series \(t\) of 4.9 and a series-cluster bootstrap interval from $+0.0006$ to $+0.0014$. Among the series where the two policies differ, 54.8 percent favor the cadence, and across 12,459 such series that majority is not close to chance. The disadvantage is small and broad. When the trigger did fire, the re-specified forecast was on average worse than simply holding the cadence form, 0.714 against 0.687 in mean MASE. At three steps, re-specifying on the validation gap chases short-window noise more often than it catches a real change.

\begin{figure}[H]
\centering
\includegraphics[width=0.98\linewidth]{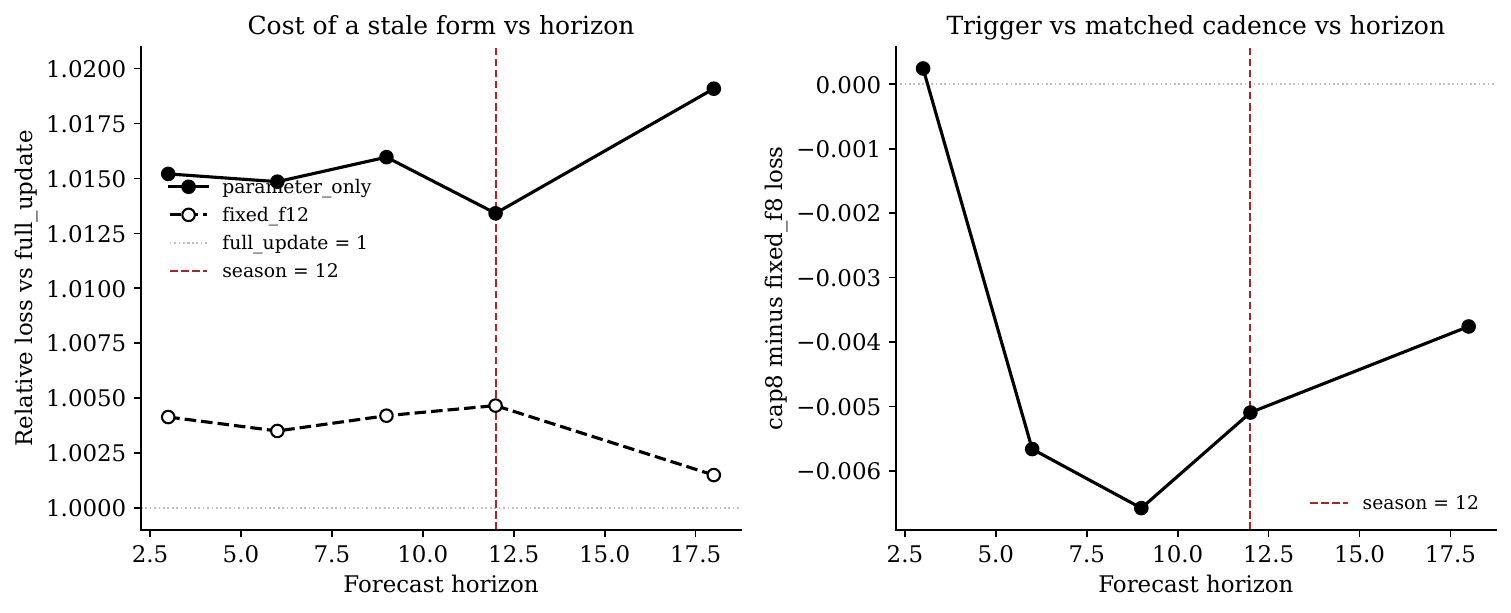}
\caption{Forecast horizon against, left, the cost of holding the model form fixed (parameter-only relative loss), which shows no clear rise across horizons, and, right, the matched comparison of the trigger against \texttt{fixed\_f8}, which turns negative beyond a three-step horizon. The vertical marker is the seasonal period.}
\label{fig:horizon}
\end{figure}

The benchmark horizon is short next to the seasonal period: three steps against twelve. Re-specification can change a model's long-range behavior, its trend and its seasonal structure, which is the extrapolation part of a forecast and not the near-term anchoring that recent data already pins down, so a short horizon is exactly where a stale form and a fresh form ought to agree. We hold the seasonal period at twelve and sweep the horizon over the fixed subsample. Two things move, and they move against each other. The cost of holding the form fixed shows no clear rise. Parameter-only relative loss reads 1.015, 1.015, 1.016, 1.013, and 1.019 across horizons 3, 6, 9, 12, and 18, a spread of six tenths of a percentage point with no step at the seasonal period. So the near-tie is not a creature of the short horizon. Changing the selected ETS form has little average effect at every depth we tried, which only sharpens the magnitude finding above. The matched comparison is the part that moves. The gap between the cap-eight trigger and \texttt{fixed\_f8} is \(+0.0002\) at horizon three and turns negative at every longer horizon, reading \(-0.0057\), \(-0.0066\), \(-0.0051\), and \(-0.0038\) at horizons 6, 9, 12, and 18. Figure \ref{fig:horizon} shows both panels.

Aggregate each series' loss difference and test across the four thousand series, and the trigger's lower average loss holds from six steps out. Table \ref{tab:horizon-paired} reports it. The per-series mean favors the trigger at every horizon beyond three, the series \(t\) runs between about minus two and minus four and a half, and the bootstrap intervals stay below zero, though the margin at eighteen is slim. The same is not true of frequency. Among the series where the two policies differ, the share favoring the cadence hovers at one half, 52 percent at six steps easing to 49 at eighteen, and none of these shares is distinguishable from one half; the per-series median difference is even positive at six and nine, where the cadence is the better forecast on the typical series. The advantage therefore lies in the size of the differences rather than their count. When the trigger helps it helps enough to move the average, and those gains outweigh a roughly equal number of small losses. A controlled experiment that varies only the timing of a form change, holding noise fixed, points the same way: the matched gap slides from a clear cadence advantage under perfectly scheduled change to a trigger advantage as the change is jittered off the schedule, crossing over near half the update interval (Appendix C.2). The sweep itself is a policy comparison rather than a pure timing decomposition, because a triggered re-specification deploys the validation winner while the cadence re-selects by AICc; the timing-isolation design varies switch timing with that selection difference held fixed across cells, so the gap's response to jitter is attributable to timing. A plausible mechanism is freshness rather than detection. When change does not arrive on a fixed cadence, gating re-specification on evidence keeps the deployed form younger than a clock does, and a longer horizon is where that shows up in the forecast.

\begin{table}[H]
\centering
\caption{Series-level comparison of \texttt{adaptive\_cap8\_tau0.8} against \texttt{fixed\_f8} by forecast horizon, on the fixed 4,000-series subsample. Each series contributes one mean loss difference, and a negative mean favors the trigger. The \(t\) is taken over the 4,000 series and the interval is a series-cluster bootstrap. Trigger worse is the share, among the \(n_{\mathrm{diff}}\) series with a nonzero mean difference, where the cadence wins; ties are excluded, and the sign-test \(p\) is against one half. None of the shares is distinguishable from one half. The three-step benchmark, where the trigger is slightly worse on the full 47,982-series run, is described in the text.}
\label{tab:horizon-paired}
\begin{tabular}{lrrrrrr}
\toprule
Horizon & Mean gap & Series \(t\) & 95\% bootstrap CI & \(n_{\mathrm{diff}}\) & Trigger worse & Sign \(p\) \\
\midrule
6 & $-0.0057$ & $-3.81$ & $[-0.0086,\,-0.0028]$ & 1,485 & 52.3\% & 0.09 \\
9 & $-0.0066$ & $-4.50$ & $[-0.0094,\,-0.0037]$ & 1,489 & 50.6\% & 0.68 \\
12 & $-0.0051$ & $-3.49$ & $[-0.0080,\,-0.0024]$ & 1,521 & 49.7\% & 0.84 \\
18 & $-0.0038$ & $-2.33$ & $[-0.0069,\,-0.0007]$ & 1,511 & 49.2\% & 0.57 \\
\bottomrule
\end{tabular}
\end{table}

The scope of the claim needs stating, because it is easy to overread. The comparison is the trigger against its own matched cadence, cap-eight against \texttt{fixed\_f8}, not against the best policy in Table \ref{tab:m4-frontier}. The claim is bounded: evidence-gated re-specification carries lower average loss than a matched fixed cadence beyond a three-step horizon, not that it is the best policy on the board and not that it wins more often. And the sweep runs on the longer series, those with enough history to reach horizon eighteen, so the result speaks to them and not to short records. Together the two panels separate the cost of staleness from the value of the evidence-gated exception. The cost of carrying a stale form barely moves with horizon. The value of departing from the clock on evidence grows once the horizon is long enough for a mistiming to surface.

\subsection{8.5 Specification-debt diagnostics}\label{specification-debt-diagnostics}

The operational trigger runs on the rolling validation score gap, but the ETS grid also supports a closed-grid diagnostic tied to the definition of specification debt. At each monitored origin under an adaptive policy, the code assigns AICc and BIC weights to the candidate forms over the full current training window and records the weight on the deployed form, giving

\[
\widetilde D^{AICc}_t = -\log w^{AICc}_t(M_t^{dep}),
\quad
\widetilde D^{BIC}_t = -\log w^{BIC}_t(M_t^{dep}).
\]

Here \(w^{\mathrm{IC}}_t(M) = \exp\{-\tfrac12(\mathrm{IC}_M - \mathrm{IC}_{\min})\} \big/ \sum_{M'} \exp\{-\tfrac12(\mathrm{IC}_{M'} - \mathrm{IC}_{\min})\}\), with the sum over the successfully fitted candidates at that origin. These are information-criterion analogues of the closed-model quantity \(D_t^{spec}\), not exact posterior model probabilities, and we compute them only as diagnostics for the finite ETS grid. The operational trigger stays the rolling validation score gap.

\begin{figure}[H]
\centering
\includegraphics[width=0.98\linewidth]{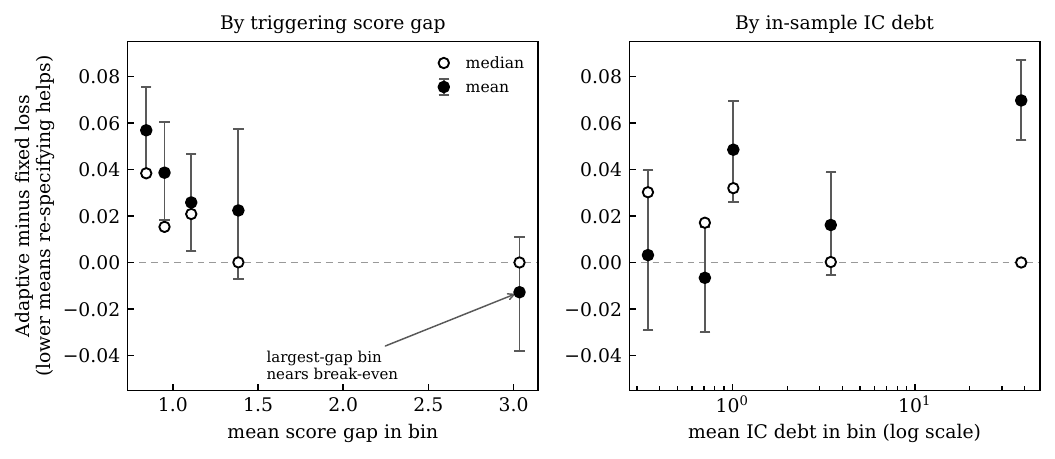}
\caption{Outcome of a triggered re-specification, binned by the signal that fired it, over all 47,982 series. Each point is a bin of triggered origins with a 95 percent series-cluster bootstrap interval on the mean (filled); the open marker is the median loss of the trigger minus \texttt{fixed\_f8}, so lower favors re-specifying. Binned by the validation score gap (left), the point estimates improve monotonically as the gap grows, from a clear loss at small gaps to break-even at the largest. Binned by the in-sample IC debt at the same origins (right, log scale), it does not: the pattern is non-monotone and the heaviest-debt bin is among the worst. The score gap ranks the outcome of acting; the in-sample criterion does not.}
\label{fig:spec-debt-bridge}
\end{figure}

The two signals are related but not the same, and only one of them tracks realized loss. Across all 47,982 series the rank association between the score gap and the IC-weight debt is weak: Spearman 0.008 for AICc weights and 0.075 for BIC weights at the matched policy, and no higher than about 0.16 for any adaptive policy in the grid. Against realized out-of-sample loss the IC debt performs worse still. A controlled test isolates the question (Appendix C.3). With the model form frozen and no selection taking place, so that the only thing at issue is whether accumulated in-sample debt anticipates degradation, the correlation between IC debt and realized loss sits near zero and changes sign with the noise level, while the out-of-sample score gap measured on the same series tracks realized loss at about 0.34 averaged over the drift cells. The in-sample criterion does not predict the quantity the trigger acts on. The same picture appears at the moment of action. Take the origins where the trigger fired and bin them by the triggering score gap. Re-specifying is costly where the gap is small, and that cost falls steadily as the gap grows: at the smallest gaps the re-specified forecast loses to the cadence by a clear margin, and only in the largest-gap bin does the difference close to break-even. The score gap ranks the outcome, running from a reliable loss at small gaps to a wash at the largest. That the ceiling here is a wash rather than a win is the three-step benchmark seen from the action side; the longer horizons of Section 8.4 are where evidence-gated action turns into lower average loss. Bin the same origins by the in-sample IC debt and no ranking appears. The pattern is ragged and non-monotone, and what little linear signal it carries runs the wrong way, with heavier debt attached to slightly worse re-specifications. Table \ref{tab:spec-debt-bridge} collects the correlations, and Figure \ref{fig:spec-debt-bridge} shows the binned outcome.

\begin{table}[H]
\centering
\caption{Specification-debt diagnostics over all 47,982 series. The first block is the rank association between the rolling score gap and the IC-weight debt at the matched policy, with the range across adaptive policies. The second is a controlled frozen-form experiment in which the model form is held fixed and only the signals are recomputed against realized loss.}
\label{tab:spec-debt-bridge}
\begin{tabular}{lr}
\toprule
Diagnostic & Value \\
\midrule
Spearman(score gap, AICc debt), \texttt{cap8\_tau0.8} & 0.008 \\
Spearman(score gap, BIC debt), \texttt{cap8\_tau0.8} & 0.075 \\
Same, range across adaptive policies & 0.003 to 0.16 \\
\addlinespace
Frozen-form drift cells: corr(score gap, realized loss) & 0.34 \\
Frozen-form drift cells: corr(IC debt, realized loss) & 0.03 \\
\bottomrule
\end{tabular}
\end{table}

The failure has a structural explanation, one that bounds the construct rather than discarding it. Information-criterion weights are in-sample quantities justified asymptotically. At 36 observations with a misspecified ETS family they reward whichever form fit the short window best after penalty, which under noise is often a form overfit to that window, so the measured debt points at overfit forms, and overfit forms forecast worse. The score-gap-to-debt association behaves the way a noise-sensitive diagnostic should: it reproduces a modest positive association under clean, longer-window estimation and frays, even turns negative, under noise. So act on the realized-loss signal, and keep the in-sample debt as a complementary diagnostic inside its valid low-noise, long-window domain. The divergence between the two is itself the finding; it refines the in-sample construct rather than refuting it.

\subsection{8.6 Cost sensitivity}\label{cost-sensitivity}

A cost-sensitive ranking should weigh computational time and instability alongside accuracy. Define

\[
C(\alpha,\gamma) = \overline{L} + \alpha\overline{T} + \gamma\overline{I},
\]

where \(\overline{L}\), \(\overline{T}\), and \(\overline{I}\) are relative loss, relative time, and relative instability. At full scale the ranking is settled by the frontier in Table \ref{tab:m4-frontier}, and Table \ref{tab:cost-winners} makes it explicit over a grid of weights. Fixed cadences take every cell up to moderate time weights: \texttt{fixed\_f4} when time is nearly free, \texttt{fixed\_f9} over most of the range, \texttt{fixed\_f12} at \(\alpha = 0.10\), and parameter-only updating takes over once computational time dominates. No adaptive policy wins a cell. The capped policies' one clear edge is instability, \texttt{adaptive\_cap12\_tau0.8} posts the lowest in Table \ref{tab:m4-frontier} at 0.930, which narrows the gap as \(\gamma\) grows but never closes it at these weights.

\begin{table}[H]
\centering
\caption{Winning policy by cost weights, \(C(\alpha,\gamma) = \bar L + \alpha \bar T + \gamma \bar I\) over all policies in the full run. Rows are the time weight \(\alpha\), columns the instability weight \(\gamma\).}
\label{tab:cost-winners}
\begin{tabular}{lllll}
\toprule
 & \(\gamma=0\) & \(\gamma=0.01\) & \(\gamma=0.05\) & \(\gamma=0.10\) \\
\midrule
\(\alpha=0\) & fixed\_f4 & fixed\_f4 & fixed\_f9 & fixed\_f9 \\
\(\alpha=0.01\) & fixed\_f9 & fixed\_f9 & fixed\_f9 & fixed\_f9 \\
\(\alpha=0.05\) & fixed\_f9 & fixed\_f9 & fixed\_f9 & fixed\_f9 \\
\(\alpha=0.10\) & fixed\_f12 & fixed\_f12 & fixed\_f12 & parameter\_only \\
\(\alpha=0.25\) & parameter\_only & parameter\_only & parameter\_only & parameter\_only \\
\(\alpha=0.50\) & parameter\_only & parameter\_only & parameter\_only & parameter\_only \\
\bottomrule
\end{tabular}
\end{table} The framework's point holds either way: model-form updating is a cost-sensitive action, not a universal calendar rule, and which policy wins depends on the loss the user actually carries.

\subsection{8.7 Interpretation}\label{interpretation}

Pull the evaluation together and it stacks into three claims of decreasing scope. Model-form search frequency has little average effect on these monthly series, at every horizon we tried, so a cheap fixed cadence is close to optimal and a trigger has little to win at the benchmark; this is the durable, full-scale version of the less-is-more result. What pays is not re-specifying as such but departing from the clock on evidence, with the timing-isolation experiment pointing to timing as the mechanism. Once the horizon is long enough for a mistimed switch to register, evidence-gated re-specification carries lower average loss than a matched fixed cadence under series-clustered tests, though it wins no more often than the clock. And the signal to act on is the out-of-sample score gap, not the in-sample information-criterion debt, which is a sound diagnostic in its domain and not a predictor of realized loss.

Underneath all of this is a distinction between two spaces. Specification debt lives in model space: a distance between the deployed form and the posterior over forms. Loss lives in forecast space. Re-specification acts in the first and is scored in the second, and the map between them is a lossy projection, not an isometry. Where the projection is collapsed, at a short horizon where parameter refitting absorbs form error and competing ETS forms give nearly identical near-term predictive distributions, on a small window where the in-sample criterion is biased toward overfit, on stable series where little changes, a real and correctly defined debt produces almost no forecast difference, acting on it cannot pay, and an in-sample estimate of it is noise. Where the projection opens up, at longer horizons and under change that does not keep to a schedule, timing begins to matter and the trigger earns its keep. The paper maps that boundary.

\section{9. Discussion}\label{discussion}

The trigger helps pull apart three decisions that often get run together.
One is whether to refresh parameters under the existing form, a decision
that can stay routine and frequent. Another is whether to re-select the
model form from an existing candidate set, which should be made
conditional on evidence and cost. The third is whether to expand the
candidate set itself by introducing new structure, and this last one
usually needs human or engineering involvement and is best reserved for
failures that recur, that can be diagnosed, and that carry real economic
weight.

The framework also sharpens how forecasting diagnostics ought to be used.
A diagnostic does not by itself call for a model change. A coverage
failure, a holiday score gap, or a segment bias is evidence, and whether
to act on it turns on the expected avoided loss set against the cost of
changing the model. This matters most in large-scale systems, where a
model can show small local failures across many series at once; without a
cost-sensitive threshold such a system tends to swing between
under-maintenance and over-maintenance.

The full-scale evaluation also sets expectations about where the rule
helps. On these monthly series the headroom is small. A cheap fixed
cadence already captures almost all of the available accuracy, so the
claim for a re-specification trigger on such data has to be narrow: it is
a timing instrument that pays once the forecast horizon is long enough for
a mistimed switch to register, not an accuracy gain at the benchmark. The
evaluation also says which signal to act on. Run the trigger on the
realized out-of-sample score gap, which tracks degradation, and not on the
in-sample information-criterion analogue of specification debt, which on
short estimation windows tilts toward forms that overfit recent noise and
does not predict realized loss. This refines the construct rather than
retiring it. In-sample debt stays a useful audit diagnostic in a
low-noise, long-window regime, and the gap between the in-sample and
out-of-sample measures is itself a reading of how far the forecast-space
projection of model-space evidence has collapsed.

A few limitations are worth stating plainly. The cost quantities are hard
to measure: computational time is observable, but analyst time,
stakeholder disruption, and the cost of instability are usually implicit,
which is exactly why a sensitivity analysis over cost ratios should be
routine. The adaptive trigger also has to be calibrated, and a badly
calibrated score threshold will over-update or under-update. Model-form
search can itself be expensive if it is implemented naively, and the
two-stage design recommended above is what protects the computational
savings of reduced updating. The benchmark is also a monthly ETS setting; the contaminated and abrupt-shift regimes that most favor
either full re-fitting or restraint are established on the simulated
designs of Appendix C rather than on M4, and the horizon result is established against a matched
cadence rather than against the full policy frontier. Finally, the
closed-form posterior threshold should not be overinterpreted in open
model spaces; it is a transparent special case, not a claim that all
production model monitoring collapses to posterior model probabilities.
The empirical evaluation drives the same point home, since the IC-weight
debt diagnostic and the rolling score gap are both informative without
being interchangeable: the IC diagnostic serves audit and interpretation
within the finite ETS grid, and the score gap is the operational trigger.

These are not defects in the decision rule so much as features of the
operational problem. Forecasting teams already make these trade-offs, just
implicitly. What this note contributes is a way to make the trade-off
explicit, auditable, and testable through rolling-origin evaluation.

\section{10. Conclusion}\label{conclusion}

Frequent full model-form updating is often unnecessary, and a forecasting
system still needs a principled way to decide when a deployed model form
has become costly enough to revisit. This note offers one. Specification
debt measures the evidence against the deployed form, the decision rule
re-specifies once expected avoidable loss exceeds re-specification cost,
and fixed update frequencies return as the special case of constant
evidence accumulation.

The full-scale evaluation then draws the boundary of the rule. On these
monthly series, model-form search frequency has little average effect at every horizon we tried, so a cheap
fixed cadence is close to optimal and the trigger does not beat it at the
benchmark. The trigger's value is real but conditional. The gain concentrates in evidence-gated departures from the
clock, posting lower average loss than a matched fixed cadence once the
forecast horizon is long enough for a mistimed re-specification to
register, with the timing-isolation experiment pointing to timing as the
operative channel, and it rests on acting
through the out-of-sample score gap rather than the in-sample
information-criterion debt, which does not predict realized loss. The
resulting picture is a practical one. Routine parameter updates can carry
on, model-form searches can happen less often, and an evidence-gated
trigger is best read as a sparse, horizon-sensitive exception sitting on
top of a strong fixed schedule, valuable exactly where the forecast-space
cost of a stale model form grows large enough to outweigh the cost of
acting.

\section{Appendix A. Derivations}\label{appendix-a.-derivations}

\subsection{A.1 Closed discrete
threshold}\label{a.1-closed-discrete-threshold}

In the closed discrete setting, write \(p_t = \pi_t(M_t^{dep})\) and
\(D_t^{spec} = -\log p_t\). The simplifying equality \(q_t = 1-p_t\)
treats every non-deployed form as actionably different from the deployed
one. In general the decision-relevant probability of actionable
misspecification is no larger than \(1-p_t\), because some alternatives can
be statistically distinct yet operationally immaterial. The decision rule
is \(q_t K c_B > c_R\). Under the strong equality
\(q_t = 1-p_t = 1-\exp(-D_t^{spec})\), substituting gives

\[
1-\exp(-D_t^{spec}) > \frac{c_R}{Kc_B}.
\]

If \(c_R < Kc_B\), then

\[
\exp(-D_t^{spec}) < 1 - \frac{c_R}{Kc_B},
\]

and hence

\[
D_t^{spec} > -\log\left(1 - \frac{c_R}{Kc_B}\right).
\]

If instead \(c_R \ge Kc_B\), the simplified threshold never fires, since
the largest avoided loss over the decision window is no greater than the
cost of re-specification. More generally, because \(q_t \le 1 - p_t\), the inequality is a necessary screening condition rather than a sufficient action rule: if it fails, re-specification cannot be justified under the binary-loss approximation, and if it holds, a calibrated estimate of actionable loss is still required.

\subsection{A.2 Fixed frequency as constant-rate
thresholding}\label{a.2-fixed-frequency-as-constant-rate-thresholding}

Write \(t_0\) for the most recent re-specification time and let
\(B_{t_0,k} = -\log\{\pi_{t_0+k}(M_{t_0}^{dep}) / \pi_{t_0}(M_{t_0}^{dep})\}\),
so that \(B_{t_0,0} = 0\) by construction. Let \(B_{t_0,k} = \lambda k\)
after a model-form update, with \(\lambda > 0\). For a threshold \(b\)
the stopping time is

\[
\tau_b = \inf\{k: \lambda k \ge b\},
\]

which gives

\[
\tau_b = \left\lceil \frac{b}{\lambda} \right\rceil.
\]

So the threshold rule coincides with a fixed update frequency when
evidence accumulates at a constant deterministic rate, and once
\(\lambda\) varies by series or by time the threshold rule turns adaptive.

\section{Appendix B. Reproducibility package}\label{appendix-b.-reproducibility-package}

The R code used for the empirical evaluation is available in the public GitHub repository \href{https://github.com/harrisonekatz/re-specification-debt}{\texttt{harrisonekatz/re-specification-debt}}. The repository contains the consolidated implementation in \texttt{R/adaptive\_update.R}, scripts for downloading the M4 monthly data, setup checks for the M4 loader and ETS backend, and scripts for running the fixed-frequency, capped adaptive, full-set, and horizon-sweep experiments reported in Section 8.

Each run writes origin-level records, policy-level summaries, diagnostics, specification-debt bridge files, triggered-series comparisons, and figures. The full-set results in Section 8 are based on the repository output folder \texttt{outputs/m4\_full\_capped}, the horizon sweep on \texttt{outputs/m4\_horizon\_h03} through \texttt{outputs/m4\_horizon\_h18}, and the series-level policy comparisons on \texttt{scripts/series\_level\_inference.R}, which reads the per-origin records for a pair of policies, aggregates each series to one mean loss difference, and reports the cross-series tests and series-cluster bootstrap intervals used in Section 8.4; the binned trigger outcomes and their intervals in Figure 3 come from \texttt{scripts/bridge\_uncertainty.R}. The IC-weight debt diagnostics are computed on the full current training window at monitored origins and are reported as audit diagnostics; the operational trigger remains the rolling validation score gap. The code is intended as an empirical evaluation of the proposed rule rather than a full reproduction of all experiments in Spiliotis and Petropoulos \citeyearpar{SpiliotisPetropoulos2024}.

\section*{Appendix C. Simulation designs and results}

Three controlled experiments support the regime, timing, and signal-validity claims of Section 8. All three run the same rolling-origin battery as the main evaluation, 36 origins, horizon three, a 36-observation training window, seasonal period twelve, on synthetic monthly series of length 120 written in the M4 format, with 600 series per cell. The policy set is a reduced battery patterned after the main run's policy classes: full updating, parameter-only updating, fixed cadences at 6, 8, and 12, pure adaptive triggers at thresholds 0.2, 0.4, 0.8, and 1.5, and capped triggers at caps 8 and 12. The scripts named below reproduce each design.

\subsection*{C.1 Shock battery}

Each series is a seasonal signal \(y_t = \ell_0\,(1 + 0.25\, s_m)\) with the base level \(\ell_0\) drawn uniformly on \([800, 1500]\) and \(s_m\) a fixed two-harmonic monthly shape, multiplied by lognormal noise with \(\sigma = 0.05\). Persistent shocks land at observation 101, inside the evaluation region. The eight cells are a control; additive outliers at rates 0.03 and 0.10 per observation, each hit multiplied by a factor drawn from \(U(2,4)\) or \(U(0.25,0.5)\) with equal probability; permanent level shifts of 0.30 and 0.80 of the base level with random sign; a six-observation transient burst of magnitude 0.60; a variance shift that triples \(\sigma\); and heavy-tailed \(t_3\) noise. Monitoring uses the main run's 12-observation window every six periods (\texttt{simulate\_shock\_series.R}, \texttt{run\_shock\_experiment.R}).

Table \ref{tab:shock-winners} gives the winning policy in each cell, relative to full updating, and Table \ref{tab:shock-matched} the matched comparison of the cap-eight trigger against \texttt{fixed\_f8}; the cap-twelve pair against \texttt{fixed\_f12} mirrors it in the repository output. Level shifts go to full updating in both severities, and no trigger setting, down to the most sensitive threshold in the battery, changes that. The contaminated cells go the other way, toward restraint. The winners there re-specify rarely in absolute terms, the pure adaptive policies at roughly 1.2 to 2.1 re-specifications per series against six for \texttt{fixed\_f6} and thirty-six for full updating, and parameter-only updating takes the transient burst outright. Re-selecting a form on a contaminated window tends to select a bad form, so firing rarely pays. The matched comparison agrees: the trigger beats its cadence under outliers, where the clock re-specifies on schedule on corrupted windows and the trigger mostly holds, and loses under level shifts, where waiting for evidence is the wrong response. In cells where the winning relative loss sits within about one percent of full updating, the control, variance-shift, and heavy-tail cells, the winner reads as a regime indicator rather than a precise ranking, and the matched comparisons in Table \ref{tab:shock-matched} carry the inferential weight.

\begin{table}[H]
\centering
\caption{Shock battery: winning policy by cell, relative loss against full updating. 600 series per cell.}
\label{tab:shock-winners}
\begin{tabular}{llll}
\toprule
Shock & Severity & Winner & Relative loss \\
\midrule
control & none & full\_update & 1.000 \\
additive outlier & rate 0.03 & adaptive\_tau0.4 & 0.893 \\
additive outlier & rate 0.10 & adaptive\_tau0.2 & 0.918 \\
level shift & 0.30 & full\_update & 1.000 \\
level shift & 0.80 & full\_update & 1.000 \\
transient burst & 0.60 & parameter\_only & 0.936 \\
variance shift & \(3\sigma\) & adaptive\_tau0.8 & 0.996 \\
heavy tail & \(t_3\) & adaptive\_tau1.5 & 0.990 \\
\bottomrule
\end{tabular}
\end{table}

\begin{table}[H]
\centering
\caption{Shock battery: paired comparison of \texttt{adaptive\_cap8\_tau0.8} against \texttt{fixed\_f8} by cell, per-series mean loss differences over 600 series. Negative favors the trigger.}
\label{tab:shock-matched}
\begin{tabular}{llrrr}
\toprule
Shock & Severity & Mean gap & SE & \(t\) \\
\midrule
control & none & $0.0000$ & 0.0001 & $-0.1$ \\
additive outlier & rate 0.03 & $-0.0646$ & 0.0071 & $-9.1$ \\
additive outlier & rate 0.10 & $-0.0489$ & 0.0055 & $-8.9$ \\
level shift & 0.30 & $+0.0624$ & 0.0028 & $+22.4$ \\
level shift & 0.80 & $+0.0109$ & 0.0010 & $+10.5$ \\
transient burst & 0.60 & $+0.0029$ & 0.0012 & $+2.4$ \\
variance shift & \(3\sigma\) & $-0.0001$ & 0.0001 & $-1.0$ \\
heavy tail & \(t_3\) & $-0.0023$ & 0.0010 & $-2.3$ \\
\bottomrule
\end{tabular}
\end{table}

\subsection*{C.2 Timing isolation}

Series alternate between two regimes: regime A carries a trended level, growth 0.006 per period, with multiplicative seasonality of amplitude 0.30, and regime B a flat level with additive seasonality of the same amplitude and a quarter-period phase shift. Estimation noise is clean throughout, \(\sigma = 0.02\), which switches off the noise-robustness channel and leaves timing as the only thing that varies. Every series carries exactly four regime switches at a base cadence of 24 observations, and the cells jitter each switch time uniformly by 0, 4, 8, or 12 observations, with a static no-switch cell as the anchor (\texttt{simulate\_timing\_isolation.R}, \texttt{run\_timing\_isolation.R}). The readout is the per-series mean loss of the cap-eight trigger minus its matched fixed-eight cadence, by cell, and Table \ref{tab:timing-jitter} gives it. The gap sits at zero on the static cell, and under perfectly regular switching the cadence wins outright, since a clock can match the schedule exactly. The cadence's advantage shrinks as the jitter grows: break-even at jitter four, which is half the trigger's eight-period update interval, clearly negative at jitter eight, and strongly negative at full irregularity, with the medians agreeing in sign throughout.

\begin{table}[H]
\centering
\caption{Timing isolation: paired comparison of \texttt{adaptive\_cap8\_tau0.8} against \texttt{fixed\_f8} by switch-timing jitter, per-series mean loss differences over 600 series per cell. Negative favors the trigger.}
\label{tab:timing-jitter}
\begin{tabular}{lrrrr}
\toprule
Cell & \(n\) & Mean gap & \(t\) & Median gap \\
\midrule
static & 600 & $0.0000$ & $-1.0$ & $0.0000$ \\
jitter 0 (regular) & 600 & $+0.0849$ & $+21.1$ & $+0.0917$ \\
jitter 4 & 600 & $+0.0017$ & $+0.3$ & $+0.0030$ \\
jitter 8 & 600 & $-0.0304$ & $-4.7$ & $-0.0279$ \\
jitter 12 (irregular) & 600 & $-0.1245$ & $-11.1$ & $-0.0766$ \\
\bottomrule
\end{tabular}
\end{table}

\subsection*{C.3 Held-form signal validity}

The drift design uses the same two-regime alternation with three switch processes, none, a regular cadence of 24 observations, and an irregular process with switch probability 0.033 per period that produces clustered, unschedulable change, crossed with clean (\(\sigma = 0.02\)) and noisy (\(\sigma = 0.12\)) estimation, 600 series per cell (\texttt{simulate\_drift\_regime.R}). For the signal-validity test the model form is selected once by AICc on the first training window and then frozen, so no selection acts on the recorded signals. At every third origin the run records three quantities for the frozen form: the six-observation validation score gap against the grid's best challenger, the AICc-weight debt, and the realized three-step test MASE (\texttt{run\_heldform\_debt.R}, \texttt{analyze\_heldform\_divergence.R}). Table \ref{tab:heldform} reports Spearman correlations across the 7,200 monitored origins in each cell.

\begin{table}[H]
\centering
\caption{Held-form correlations with realized loss by drift cell, Spearman over 7,200 monitored origins per cell.}
\label{tab:heldform}
\begin{tabular}{llrr}
\toprule
Drift & Noise & Score gap vs loss & IC debt vs loss \\
\midrule
irregular & clean & 0.595 & $+0.253$ \\
irregular & noisy & 0.342 & $-0.085$ \\
regular & clean & 0.205 & $-0.077$ \\
regular & noisy & 0.227 & $+0.020$ \\
static & clean & 0.044 & $+0.022$ \\
static & noisy & 0.408 & $-0.354$ \\
\bottomrule
\end{tabular}
\end{table}

The score gap is positive in every cell and holds its sign under noise. Averaged over the four drift cells it sits at 0.34 against the IC debt's 0.03, and the IC debt flips sign with the noise level, reaching \(-0.354\) in the static noisy cell. The claim should stay comparative rather than absolute. The static noisy cell puts the score gap at 0.408 with no drift present, a shared-badness floor: a misfit form is bad on the validation tail and on the test window alike, and noise inflates both. The defensible statement is the contrast, that the realized-loss signal tracks degradation far more stably than the in-sample surrogate computed at the same origins, not that the score gap is a clean drift detector.

\renewcommand\refname{References}
  \bibliography{references.bib}

\end{document}